# Resolution-Based Distillation for Efficient Histology Image Classification

Joseph DiPalma, BS[1], Arief A. Suriawinata, MD[2], Laura J. Tafe, MD[2],
Lorenzo Torresani, PhD[1], Saeed Hassanpour, PhD[1,3,4*]

[1]Department of Computer Science, Dartmouth College, Hanover, NH 03755, USA

[2]Department of Pathology and Laboratory Medicine, Dartmouth-Hitchcock Medical Center, Lebanon, NH 03756, USA

[3]Department of Biomedical Data Science, Geisel School of Medicine at Dartmouth, Hanover, NH 03755, USA

[4]Department of Epidemiology, Geisel School of Medicine at Dartmouth, Hanover, NH 03755, USA

[*]Corresponding Author: Saeed Hassanpour, PhD
Postal address: One Medical Center Drive, HB 7261, Lebanon, NH 03756, USA
Email: Saeed.Hassanpour@dartmouth.edu

# Abstract

Developing deep learning models to analyze histology images has been computationally challenging, as the massive size of the images causes excessive strain on all parts of the computing pipeline. This paper proposes a novel deep learning-based methodology for improving the computational efficiency of histology image classification. The proposed approach is robust when used with images that have reduced input resolution and it can be trained effectively with limited labeled data. Pre-trained on the original high-resolution images, our method uses knowledge distillation to transfer learned knowledge from a teacher model to a student model trained on the same images at a considerably lower resolution. Also, to address the lack of large-scale labeled histology image datasets, we perform the knowledge distillation in a self-supervised fashion. We evaluate our approach on two distinct histology image datasets associated with celiac disease and lung adenocarcinoma. Our results on both datasets demonstrate that a combination of knowledge distillation and self-supervision allows the student model to approach, and in some cases, surpass the classification accuracy of the teacher model, while being much more computationally efficient. Additionally, we observe an increase in student classification performance as the size of the unlabeled dataset increases, indicating that there is potential for this method to scale further with additional unlabeled data. For celiac disease, our model outperforms the high-resolution teacher model in terms of accuracy, F1-score, precision, and recall, while requiring 4 times fewer computations. For lung adenocarcinoma, our results at 1.25x magnification are within 3% of the results for the teacher model at 10x magnification, with a reduction in computational cost by a factor of 64. Furthermore, our celiac disease outcomes benefit from additional performance scaling with the use of more unlabeled data. In the case of 0.625x magnification, using unlabeled data improves accuracy by 4% over the baseline.

Therefore, our approach can improve the feasibility of using deep learning solutions for digital pathology based on standard computational hardware and infrastructures.

# 1. INTRODUCTION

Digital pathology was introduced over 20 years ago to facilitate viewing and examining high-resolution scans of histology slides. A digital scanning process produces whole-slide images (WSIs), which are then analyzed with computational tools [1,2]. While digital scans circumvent traditional microscope use, they introduce new computational challenges. The resulting WSIs can be as large as 150,000×150,000 pixels in size and occupy gigabytes of space. Software like OpenSlide [3] and QuPath [4] have been instrumental in providing tools to read and analyze WSIs. Nevertheless, these tools are only a single step in the pipeline to store and analyze WSIs, and they cannot address other computational requirements such as storage capacity, network bandwidth, computing power, and graphics processing unit (GPU) memory.

In recent years, computer vision-based deep learning methods have been developed for digital pathology [5–7]; however, their application and scope have been limited due to the massive size of WSIs. Figure 1 illustrates the magnitude of a sample histology image. Even with the most recent computational advancements, deep learning models for analyzing WSIs are still not feasible to run on all except the most expensive hardware and GPUs. These computational constraints for analyzing high-resolution WSIs have limited the adoption of deep learning solutions in digital pathology.

This paper addresses this computational bottleneck through implementing a deep learning approach designed to operate accurately on lower-resolution versions of WSIs. This approach aims to lower the resolution of the input image while minimizing its effect on the classification performance. By operating on WSIs with a lower resolution, our approach allows for slides to be scanned at a lower resolution, reducing scanning time and the strain on computational hardware and infrastructure.

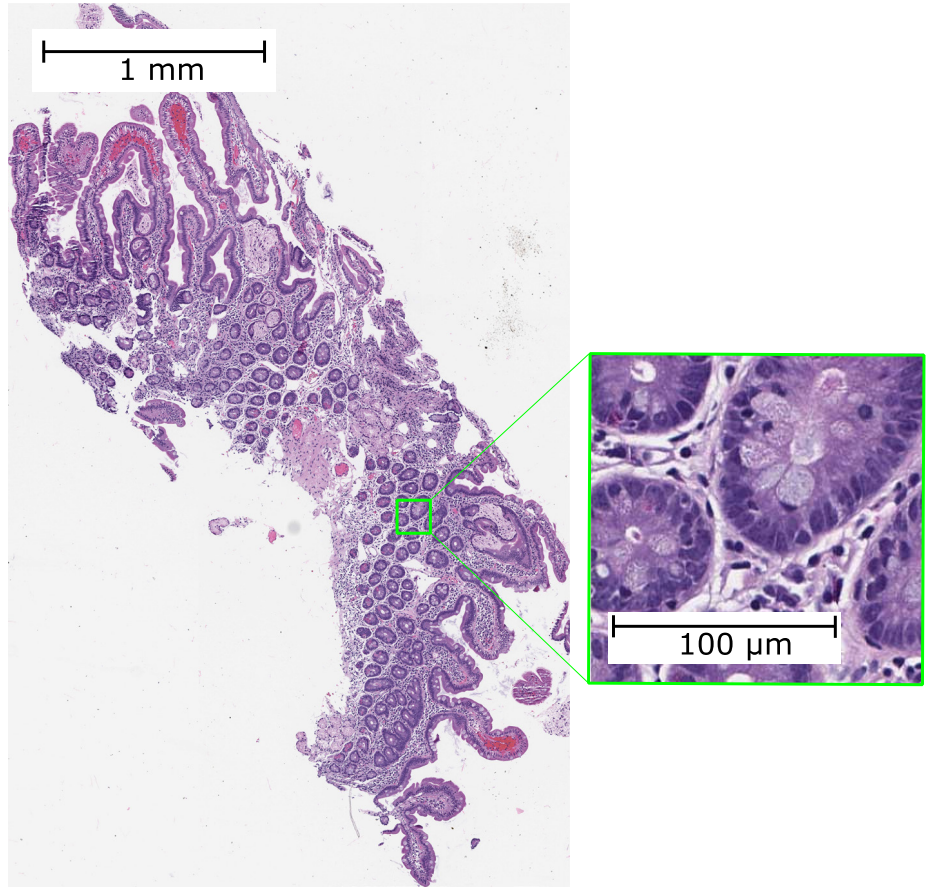

**Figure 1.** A sample WSI intended to show the high resolution and large size of histology images.

Our proposed methodology is a novel approach to make high-resolution histology image analysis more efficient and feasible on standard hardware and infrastructure. Specifically, we propose a knowledge distillation-based method where a teacher model works at a high resolution and a student model operates at a low resolution. We aim to distill the teacher model's learned representation knowledge into the student model that is trained at a much lower resolution. The knowledge distillation is performed in a self-supervised fashion on a larger unlabeled dataset from the same domain. Large, labeled datasets are hard to find in the medical field, leading us to adopt a self-supervised approach to account for the lack of access to sizeable, labeled histology image datasets. This knowledge distillation method can potentially increase the model's performance on lower-resolution images, while simultaneously saving significant amounts of memory and computation.

## 2. RELATED WORK

### 2.1 Histology Image Classification

Previously, several methods have been proposed to solve the WSI classification problem. Some approaches work by tiling the WSI into more reasonably sized patches and learning to classify at the patch level [5–8]. In some recent works, the patch level predictions are aggregated using simple heuristic rules to produce a slide level prediction [5,6,8]. These rules are modeled after how pathologists classify WSIs in clinical practice. In another work, a simple maximum function was used on patch-based slide heat maps for whole-slide predictions [7]. While these methods achieved reasonable overall performance, their analyses are fragmented, and they do not incorporate the relevant spatial information into the training process. We aim to avoid patch-based processing since it introduces additional computational overhead that can be bypassed with tissue- or slide-based analysis methods.

Multiple-instance learning (MIL) has been proposed to address the slide-level labeling problem [9–13]. MIL is a supervised learning scheme where data-points, or instances, are grouped into bags. Each bag is labeled with the class by the instance count of that particular class. MIL is well-suited towards histology slide classification, as it is designed to operate on weakly-labeled data. MIL-based methods better account for the weakly-labeled nature of patches, but they still tend to miss the holistic slide information.

Recent work has shown that operating at the slide-level is possible by splitting up the computation into discrete units that can be run on commodity hardware [14,15]. The overall calculation is equivalent to the one performed at the slide-level due to the invariance of most layers in a convolutional neural network. This method aims to analyze WSIs at the original high-resolution level to avoid losing larger context and fine details. Although this approach helps run large neural networks, it still requires considerable computational resources to analyze WSIs at a high resolution.

Attention-based processes have also been suggested for WSI analysis. Attention-based mechanisms divide the high-resolution image into large tiles and simultaneously learn the most critical regions of WSIs for each class and their labels [16–18]. Although these methods achieve high classification performance, they still necessitate considerable computational resources to analyze high-resolution images.

## 2.2 Self-Supervised Learning

Self-supervised learning is a machine learning scheme that allows models to learn without explicit labels. Large, unlabeled datasets are readily accessible in most domains, and self-supervised methods can assist in improving classification performance without requiring resource-intensive, manually labeled data. In this scheme, learning occurs using a pre-text task on an inherent attribute of the data. As the pre-text task operates on an existing data feature, it requires no manual intervention and can be easily scaled. Proposed pre-text tasks include colorization [19,20], rotation [21,22], jigsaw puzzle [23], and counting [24]. Recent studies have explored the invariance of histology images to affine transformations, but none use self-supervised learning [25,26]. Several other works have proposed self-supervised techniques for histology images exploiting domain-specific pre-text tasks, including slide magnification prediction [27], nuclei segmentation [28], and spatial continuity [29]. In contrast, our work introduces a new pre-text task designed to transfer the knowledge present in models trained on high-resolution WSIs to ones operating on low-resolution WSIs.

## 2.3 Knowledge Distillation

Knowledge distillation has proven to be a valuable technique for transferring learned information between distinct models with different capacities [30,31]. As models and datasets exponentially increase in size, it is critical to adapt our methods accordingly to support less powerful devices [32]. Knowledge distillation has been beneficial to many areas of computer vision such as semantic segmentation [33], facial recognition [34,35], object detection [36], and classification [37]. Although some prior work has used knowledge distillation for chest X-rays in the medical domain [38], knowledge distillation has not been widely used for histology image analysis.

Initial knowledge distillation studies used neural network output activations, called logits, to transfer the learned knowledge from a teacher model to a student model [30–32]. FitNet built upon this knowledge distillation paradigm by suggesting that while the logits are important, the intermediate activations also encode the model's knowledge [39]. This method proposed adding a regression term to the knowledge distillation objective to improve the overall performance of the student model while reducing the number of parameters. In this paper, we model our architecture after the FitNet approach to maintain the spatial correspondence between teacher and student models, as it represents clinically relevant information. Of note, in contrast to our approach, previous work in this domain does not include self-supervision [40]. As we show later in this paper, self-supervision proves to be a deciding factor in increasing overall classification performance for histology images.

## 3. TECHNICAL APPROACH

### 3.1 Overview

There are two main phases and one optional phase to our approach as follows:

1. Train a teacher model at 10x magnification on the labeled dataset, as explained in Section 3.2.
2. Train the knowledge distillation model on the unlabeled dataset from 10x to a lower magnification, explained in Section 3.3 and shown in Figure 2.
3. (Optional) Fine-tune the student model using the labeled dataset at a lower resolution, as explained in Section 3.4.

### 3.2 Teacher Model

For the teacher model, we used a residual network (ResNet) [41]. ResNet was chosen due to its excellent empirical performance compared to other deep learning architectures. We used the built-in ResNet PyTorch implementation [42].

The teacher model input was slides at 10x magnification (1 μm/pixel), which are considered high-resolution. We performed online data augmentation consisting of random perturbations to the color brightness, contrast, hue, and saturation, horizontal and vertical flips, and rotations. Additionally, each input was standardized by the mean and standard deviation of the training set across each color channel.

### 3.3 Knowledge Distillation from High-Resolution

Knowledge distillation (also referred to as 'KD') is a machine learning method, where typically a larger, more complex model "teaches" a smaller, simpler student model what to learn [30]. The learning occurs by optimizing over some desired commonality between the models. For our approach, we opted to keep the student and teacher model architectures identical and instead modified the input resolution. As input data resolution is a significant factor for efficient and accurate histology image analysis, we decided that the teacher model should receive the original high-resolution image as input while the student model receives a low-resolution input. For optimizing our knowledge distillation model, the total loss is the sum of (1) the soft loss and (2) the pixel map. These loss components are described below, and an overview of our knowledge distillation approach is shown in Figure 2.

$$Loss_{total} = Loss_{soft} + Loss_{pixel} \qquad (1)$$

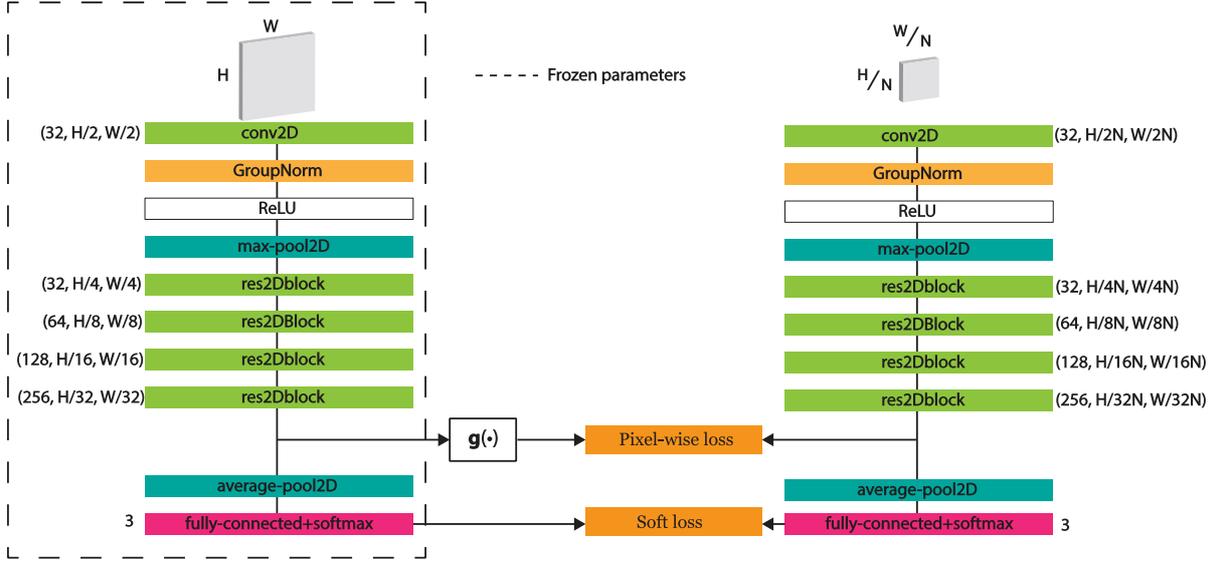

**Figure 2.** Overview of the knowledge distillation model. The **g**(·) block is a resizing function that scales the teacher feature maps to the same size as the corresponding student ones. The Pixel-wise and Soft losses are combined to produce the total loss for the optimization process.

To promote classification similarity between the teacher and student models, we utilized the Kullback-Leibler (KL) Divergence over the outputs of the teacher and student models as the loss function [30,43]. Additionally, the loss function is "softened" by adding a temperature $T$ to the softmax computation. Intuitively, softening the loss function gives more weight to smaller outputs, thus transferring information that would have been overpowered by greater values. The soft loss is computed as follows:

$$Loss_{soft} = KL\left(\sigma\left(\frac{\overrightarrow{F_{fc}^t}}{T}\right), \sigma\left(\frac{\overrightarrow{F_{fc}^s}}{T}\right)\right) \cdot T^2 \qquad (2)$$

where $\overrightarrow{F_{fc}^t}, \overrightarrow{F_{fc}^s}$, and $\sigma(\cdot)$ represent the teacher logits, student logits, and softmax function, respectively. Note that we multiply by $T^2$ since the gradients will scale inversely to this factor [30].

To ensure that the teacher and student models focus on similar areas, we compute the mean-squared error over the feature map outputs. We introduce $\mathbf{g_1}(\cdot)$ and $\mathbf{g_2}(\cdot)$ as the max-pooling and bicubic interpolation operations, respectively. We use both max-pooling and bicubic interpolation for the pixel-wise loss. As shown in the Supplementary Material, we found these two functions provide the most consistent performance for the loss function when combined. The pixel-wise loss is computed as follows:

$$Loss_{pixel} = \frac{1}{2} \cdot \sum_{k=1}^{2} \|\mathbf{g_k}(\mathbf{F}_{conv}^{t}) - \mathbf{F}_{conv}^{s}\|^2 \qquad (3)$$

where $\mathbf{F}_{conv}^{t}$ and $\mathbf{F}_{conv}^{s}$ are the outputs of the final convolutional layer in the teacher and student models, respectively. We require the functions $\mathbf{g_1}(\cdot)$ and $\mathbf{g_2}(\cdot)$ since the size of the teacher feature map outputs are $N^2$ times larger than the student ones, ignoring negligible differences due to rounded non-integer dimensions in some instances.

### 3.4 Fine-Tuning

After performing the knowledge distillation, we fine-tuned the student model weights on the lower resolution training dataset. The goal of fine-tuning the model is to make small weight adjustments for maximal performance on the labeled data without undoing the learning in the previous layers. To this end, all weights were frozen except the ones in the fully connected layer. The weights were trained using the Adam optimization algorithm [44] until convergence. The data augmentations enumerated in Section 3.2 were applied to the input data. Similar to the teacher model training, we used the cross-entropy loss to learn ground-truth labels in this phase. We opted to skip this phase in experiments where the same training set was used across Phases I and II, as it resulted in lower classification performance on the validation set due to overfitting on the training set.

### 3.5 Gradient Accumulation

We used gradient accumulation to account for the large and variable sizes of the slides. Gradient accumulation computes the forward and backward pass changes for each input, but it does not update the model weights until all mini-batch backward passes are complete. While gradient accumulation does not affect most layers, batch normalization layers are affected, as they require operation on a mini-batch to work correctly. In our model, instead of batch normalization, we used group normalization, which has more consistent performance across varying mini-batch sizes [45] due to its independence from the mini-batch dimension. This modification allows the model to learn properly with gradient accumulation.

## 4. EXPERIMENTAL SETUP

### 4.1 Datasets

We perform our experiments on two independent datasets collected at the Dartmouth-Hitchcock Medical Center, a tertiary academic medical center in New Hampshire, USA. The slides were hematoxylin-eosin stained formalin-fixed paraffin-embedded and digitized by a Leica Aperio scanner at either 20x or 40x magnification. Every downsampling was obtained directly from the original image to avoid any potential artifacts caused by a composition of downsamplings. To generate the required low-resolution WSIs, we used the Lanczos filter to create several downsampled versions of each image [46]. We used the notation nx magnification relative to the original magnification. For example, an originally 20x slide downsampled four times in both height and width dimensions would have $n = 20/4 = 5$ and be denoted 5x.

This study and the use of human participant data in this project were approved by the Dartmouth-Hitchcock Health Institutional Review Board (IRB) with a waiver of informed consent. The conducted research reported in this article is in accordance with this approved

Dartmouth-Hitchcock Health IRB protocol and the World Medical Association Declaration of Helsinki on Ethical Principles for Medical Research involving Human Subjects.

**4.2 Celiac Disease Dataset**

Celiac disease (CD) is a disorder that is estimated to impact 1% of the population worldwide [47,48]. Diagnosing and treating CD is clinically significant, as undiagnosed CD is associated with a higher risk of death [47,48]. A duodenal biopsy is considered the gold standard for CD diagnosis [49]. A pathologist examines these biopsies under a microscope to identify the markers associated with CD.

Our CD dataset is from 1,364 patients distributed across the Normal, Non-specific Duodenitis, and Celiac Sprue classes. Each patient has one or more WSIs consisting of one or more tissues. A gastrointestinal pathologist diagnosed every slide as either Normal, Non-specific Duodenitis, or Celiac Sprue.

The CD slides contained significant amounts of background. Hence, as a preprocessing step, we used the tissueloc [50] code repository to find approximate bounding boxes around the relevant regions of the slide using a combination of image morphological operations. This process aids in reducing the computational burden while simultaneously removing the clinically unimportant background regions.

We partitioned the dataset into a labeled set and an unlabeled auxiliary set. The auxiliary dataset (AD) is obtained by ignoring the labels. Our labeled set comprises 300 patients distributed uniformly across the Normal, Non-specific Duodenitis, and Celiac Sprue classes. A 70% training, 15% validation, and 15% testing split was produced by randomly partitioning the patients. In Table 1, we show the tissue counts for all sets.

|  | Supervised | | | Self-Supervised | | |
| --- | --- | --- | --- | --- | --- | --- |
| Class | Training | Validation | Testing | ADv1 | ADv2 | Development |
| Normal | 1,182 | 253 | 241 | 4,774 | 16,661 | 441 |
| Non-specific Duodenitis | 2,202 | 390 | 469 | 130 | 265 | 583 |
| Celiac Sprue | 2,524 | 524 | 529 | 416 | 1,799 | 921 |

**Table 1.** Distribution of the CD tissues for all sets used in the model. The class counts for the self-supervised sets ADv1 and ADv2 are only provided as a reference and this class information was not used in the self-supervision process.

For self-supervision, we randomly sampled from the CD slides not used in any of the training, validation, or testing sets. In order to explore the effects of unlabeled dataset size, we created two auxiliary datasets, ADv1 and ADv2, such that ADv1⊂ADv2. ADv1 and ADv2 comprise 300 and 1,004 patients, respectively. Experimenting with two unlabeled datasets allowed us to demonstrate the efficacy of our method as the dataset size scales. We also sampled an additional 20 patients from each class to use as a proxy development set for hyperparameter tuning. The 60-patient development set was intended to validate the self-supervision process and remained independent from the test set used for evaluation. The distribution for these datasets for self-supervised learning is shown in Table 1.

**4.3 Lung Adenocarcinoma Dataset**

Lung cancer is the leading cause of cancer death in the United States [51]. Of all histologic subtypes, lung adenocarcinoma (LUAD) is the most common pattern [52] and its rates continue to increase among certain subpopulations [53]. The World Health Organization identifies five predominant subtypes: lepidic, acinar, papillary, solid, and micropapillary for lung adenocarcinoma, in order of increasing severity [54]. The classification of lung adenocarcinoma subtypes on histology slides has proven to be especially challenging, as over 80% of cases contain mixtures of multiple patterns [55,56].

| Class | Training | Testing |
|---|---|---|
| Lepidic | 514 | 81 |
| Acinar | 691 | 124 |
| Papillary | 43 | 9 |
| Micropapillary | 411 | 55 |
| Solid | 424 | 36 |

**Table 2.** Distribution of the LUAD tissues for all sets used in the model. The counts correspond to the annotations provided by the pathologists.

Our LUAD dataset was randomly split into two sets, with 235 slides for training and 34 slides for testing. Both the training and testing sets were annotated by pathologists for predominant subtypes. As such, each slide in the training and testing set consists of at least one tissue. Some training and testing slides contained tissues annotated as benign, but we removed them as they are trivial to identify and are not related to cancer subtypes. Given the considerably smaller size of this dataset compared to the CD dataset, we did not perform any experiments on varying unlabeled dataset sizes and used the entire training set for all runs. No hyperparameter tuning was performed for this model, and we used the same configuration as the CD equivalent. The distribution of the LUAD data is presented in Table 2 for both training and testing sets.

**4.4 Implementation Details**

We evaluated all models on the labeled test set corresponding to each training dataset. No data augmentation was applied to the test sets beyond standardizing the color channels by the mean and standard deviation of the respective labeled training sets. To evaluate our classification performance, we used accuracy, F1-score, precision, and recall. These metrics were computed in a one-vs.-rest fashion for each class. We computed the mean value for each metric by macro-averaging over all classes. The 95% confidence intervals (CIs) were produced using bootstrapping on the test set for 10,000 iterations. We calculate each model's computational cost by counting the number of billions of floating-point operations (GFLOPS) for a forward pass of that model. Using the number of GFLOPS allows us to evaluate the performance gains while also considering the computational cost.

**Teacher Model.** We trained the teacher model on high-resolution input images at 10x magnification. The He initialization scheme [57] was used to initialize the weights. We utilized the Adam optimization algorithm [44] for 100 epochs of training with a learning rate of 0.001. The Adam optimizer minimized the cross-entropy loss function with respect to the ground-truth slide labels.

**Baseline.** All baseline models are trained on a specified magnification from randomly initialized weights using the He initialization scheme [57]. We use the same ResNet architecture as the teacher model for these baselines.

**KD.** Our knowledge distillation approach consists of a teacher model, described above, and a student model of the same ResNet architecture. In contrast to the standard ResNet architecture, we use both the final convolutional and fully connected layer outputs as our unlabeled hints and feature recognition knowledge, respectively. We use the labeled training and validation sets for the distillation and ignore the labels in the self-supervised part of our approach. As explained in Section 3.4, we do not apply fine-tuning for these experiments as it contributes to overfitting according to our validation set.

**KD (AD).** The knowledge distillation approach using the auxiliary datasets in this paper is similar to stock distillation [30]. The main difference is that we utilized unlabeled auxiliary datasets for self-supervised learning instead of using a labeled dataset.

# 5. RESULTS

In Table 3, we present the results of the teacher model trained from scratch at 10x magnification for the CD and LUAD test sets.

|  | CD | LUAD |
|---|---|---|
| Accuracy | 87.06 (85.65-88.48) | 92.59 (90.11-94.94) |
| F1-Score | 75.44 (72.31-78.51) | 84.81 (79.09-90.07) |
| Precision | 75.62 (72.55-78.66) | 84.62 (78.49-90.31) |
| Recall | 77.15 (74.19-80.06) | 85.31 (79.27-90.66) |

**Table 3.** Results and the corresponding 95% CIs for the teacher model trained at 10x magnification as percentages. The above results were obtained on the respective test sets, detailed in Sections 4.2 and 4.3.

We present the results of our proposed approach for all tested magnifications in Table 4. The performance and computational costs of our models are shown in Figure 3.

| | Celiac Disease | | | | Lung Adenocarcinoma | |
|---|---|---|---|---|---|---|
| | Baseline | KD | KD (ADv1) | KD (ADv2) | Baseline | KD |
| **mag = 0.625x** | | | | | | |
| Accuracy | 79.11 (77.74-80.54) | 81.31 (79.91-82.72) | 82.55 (81.15-83.97) | **83.17 (81.75-84.61)** | 84.16 (81.50-86.87) | **85.60 (82.88-88.37)** |
| F1-Score | 55.72 (52.08-59.34) | 64.16 (60.92-67.37) | 64.95 (61.43-68.45) | **66.83 (63.46-70.14)** | 64.45 (56.79-71.74) | **68.03 (60.49-75.12)** |
| Precision | 56.20 (52.40-59.96) | 64.27 (60.91-67.56) | 66.65 (62.94-70.32) | **67.21 (63.78-70.52)** | 66.48 (58.60-74.03) | **68.04 (60.47-75.39)** |
| Recall | 55.67 (52.17-59.16) | 65.13 (62.06-68.20) | 64.11 (60.64-67.61) | **69.29 (66.06-72.42)** | 63.60 (55.85-71.24) | **68.49 (60.78-75.82)** |
| **mag = 1.25x** | | | | | | |
| Accuracy | 82.70 (81.23-84.14) | 84.03 (82.61-85.47) | 84.43 (83.01-85.85) | **84.87 (83.40-86.32)** | 86.79 (84.04-89.62) | **90.80 (88.24-93.32)** |
| F1-Score | 65.06 (61.55-68.43) | 70.49 (67.45-73.55) | 69.75 (66.53-72.94) | **71.20 (67.89-74.40)** | 70.39 (63.21-77.38) | **80.33 (74.02-86.15)** |
| Precision | 65.06 (61.51-68.48) | 70.53 (67.39-73.66) | 69.32 (66.13-72.51) | **71.14 (67.81-74.35)** | 71.62 (64.39-78.56) | **79.79 (73.54-85.87)** |
| Recall | 65.22 (61.70-68.63) | 71.06 (68.10-74.02) | 70.95 (67.72-74.17) | **73.56 (70.40-76.61)** | 70.25 (62.75-77.63) | **81.77 (75.41-87.69)** |
| **mag = 2.5x** | | | | | | |
| Accuracy | 83.71 (82.29-85.17) | 85.68 (84.25-87.13) | 85.38 (83.94-86.78) | **85.83 (84.38-87.27)** | 91.39 (88.89-93.76) | **91.40 (88.84-93.90)** |
| F1-Score | 68.32 (64.92-71.66) | 73.01 (69.94-76.03) | 72.39 (69.21-75.41) | **73.56 (70.42-76.64)** | **82.15 (76.15-87.60)** | 81.22 (75.03-87.04) |
| Precision | 68.23 (64.77-71.67) | **74.74 (71.57-77.90)** | 72.99 (69.76-76.06) | 73.61 (70.44-76.68) | **83.61 (77.47-89.33)** | 81.27 (75.26-86.92) |
| Recall | 68.57 (65.13-71.98) | 74.67 (71.99-77.28) | 75.67 (72.86-78.34) | **76.43 (73.61-79.17)** | 81.50 (75.06-87.45) | **83.23 (77.07-88.85)** |
| **mag = 5x** | | | | | | |
| Accuracy | 86.15 (84.71-87.61) | 85.74 (84.28-87.21) | 86.99 (85.54-88.46) | **87.20 (85.78-88.62)** | 92.09 (89.61-94.50) | 92.00 (89.43-94.36) |
| F1-Score | 73.42 (70.15-76.63) | 73.27 (70.19-76.33) | 75.07 (71.89-78.17) | **75.86 (72.71-78.92)** | **83.61 (77.69-89.05)** | 83.31 (77.19-88.79) |
| Precision | 73.44 (70.12-76.68) | 75.10 (71.91-78.23) | **76.46 (73.42-79.44)** | 76.07 (72.95-79.13) | 84.74 (78.78-90.31) | 84.11 (78.29-89.33) |
| Recall | 73.65 (70.41-76.93) | 74.82 (72.16-77.51) | **78.00 (75.18-80.72)** | 77.41 (74.41-80.35) | 83.09 (76.87-88.87) | **84.93 (79.33-90.05)** |

**Table 4.** Results for baseline and KD approaches as percentages with corresponding 95% CIs. Baseline models were trained from scratch until convergence on the corresponding magnification. The KD model without an auxiliary dataset was trained using the labeled dataset. **Boldface** text indicates the best performing model for each magnification and metric.

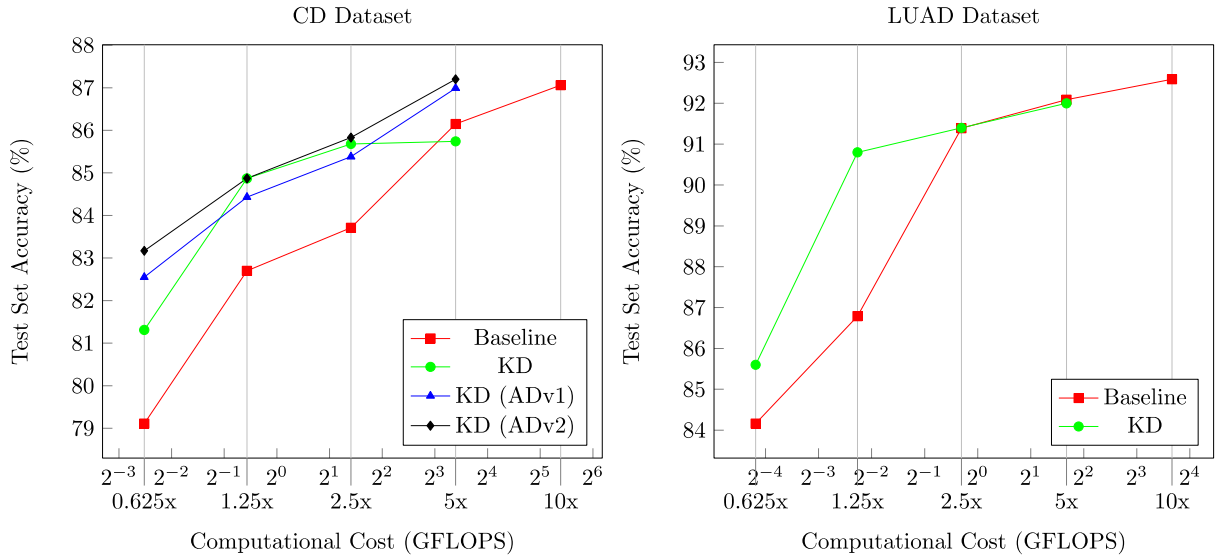

**Figure 3.** Test set accuracy plotted against the computational cost. The computational cost is measured in GFLOPS and corresponds to the approximate number of floating-point operations per forward pass. The magnification of the model input data is displayed under the computational cost values.

## 6. DISCUSSION

As presented in Table 4, our KD method outperforms the baseline metrics in all trials for celiac disease. The lung adenocarcinoma results show that our approach improves performance for 0.625x (16 μm/pixel), 1.25x (8 μm/pixel), and 2.5x (4 μm/pixel), and is close to the baseline performance, with a slight difference, for 5x (2 μm/pixel) input images. This outcome is consistent with our 5x results on the CD dataset without the AD self-supervision phase.

While adding more data helped to increase CD classification accuracy at 0.625x magnification by over 4%, this performance benefit narrowed as the magnification increased further. This trend can be seen in Figure 3, where the test set accuracy curves approach each other as the computational cost grows. Most importantly, our method outperforms the baseline at 10x magnification for the distillation approaches on the auxiliary dataset. This performance gain comes with at least a 4x reduction in computational cost.

Using our model to maintain accurate classification performance while minimizing computational cost could facilitate scanning histology slides at a much lower resolution. According to the Digital Pathology Association, scanners cost up to $300,000 depending on the configuration [58]. Reducing the scanning resolution would have a two-fold benefit, lessening both the scan time and scanner cost. Beyond the scanner, storing and analyzing lower resolution WSIs would be less burdensome on the computational infrastructure. Instead of investing in complex data solutions, pathology laboratories could migrate to cloud-based services that could manage and analyze smaller datasets using standard network bandwidth.

There are still some improvement areas for our work, namely evaluating our model on additional datasets from different institutions. While our method was validated on two datasets, both are collected from our institution and may contain inherent biases in staining and slide preparation. With more datasets, we would be able to investigate the scaling effects of self-supervised learning beyond the size of our existing dataset. The impact of scaling could prove especially useful for smaller healthcare facilities that may not have the capabilities to collect and label data as required for training a typical deep learning model for histology image analysis. In addition to larger datasets, it is important to explore the efficacy of this methodology on slides from different medical centers and various diseases to evaluate the generalizability of our proposed approach.

Although the trained models can be used on WSIs with lower resolutions, our method still requires using high-resolution WSIs during training. While reducing the computational requirements of the inference stage is always beneficial, there is no reduction in cost for training the teacher model or the self-supervised and knowledge-distillation models. This weakness is an active area of investigation in our future work. One possibility is using transfer learning to adapt a pre-trained model to an alternative high-resolution histology dataset. A method that utilizes transfer learning in this fashion would remove the burden of continuously retraining teacher models for each new dataset. Lastly, we plan to combine our approach with a neural network visualization method. As pathologists rely on visual markers to diagnose slides, it is critical to provide humanly interpretable visualizations and insights to avoid a black-box approach in histology image analysis.

## 7. CONCLUSION

In this work, we demonstrated that knowledge distillation could be applied to histology image analysis and further improved by self-supervision. We showed that our method both improves performance at significantly lower computational cost and scales with dataset size. The empirical evidence presented proves that it is possible to transfer information learned across magnifications and still produce clinically meaningful results. Our approach allows for scanning WSIs at a significantly lower resolution while having little to no classification accuracy degradation. Our method removes a major computational bottleneck in the use of deep learning for histology image analysis and opens new opportunities for this technology to be integrated into the pathology workflow.

**ACKNOWLEDGMENTS**

The authors would like to thank Lamar Moss and Behnaz Abdollahi for their help and suggestions to improve the manuscript. We would also like to thank Naofumi Tomita for his feedback on the draft and help in producing figures.

**CONFLICT OF INTEREST STATEMENT**

None Declared.

**FUNDING**

This research was supported in part by grants from the US National Library of Medicine (R01LM012837) and the US National Cancer Institute (R01CA249758).

# SUPPLEMENTARY MATERIAL

**Ablation Study**

We performed an ablation study on the resizing function, $g(\cdot)$, used to match sizes between the teacher and student models in the knowledge distillation (KD) phase. This evaluation was designed to study the empirical performance of max-pooling (MP), bicubic interpolation (INT), and their combinations. Note that the CD dataset was used for all ablation study experiments. Specifically, we tried the following configurations:

1. **KD.** Knowledge Distillation with only soft loss on the teacher and student logits, i.e., no resizing function $g(\cdot)$ since the intermediate convolutional activations are unused.

2. **KD+MP.** Knowledge Distillation with soft loss on teacher and student logits as well as pixel-wise loss on teacher outputs resized using max-pooling.

3. **KD+INT.** Knowledge Distillation with soft loss on teacher and student logits as well as pixel-wise loss on teacher outputs resized using bicubic interpolation.

4. **KD+MP+INT.** Knowledge Distillation with soft loss on teacher and student logits as well as pixel-wise loss on teacher outputs resized using both max-pooling and bicubic interpolation. We averaged the loss value produced for each resizing operation.

We performed this ablation study using the same training procedure noted in Section 4.4. The results of our ablation study are shown in Supplementary Figure 1. We chose to use KD+MP+INT in our self-supervised experiments, as it outperformed the other methods in most cases and remained consistent across magnifications. The combination of max-pooling and bicubic interpolation probably helped to smooth out any noise in the self-supervision process by reducing the inaccuracies introduced by resizing the feature maps. Notably, no one resizing scheme outperformed the others in all cases. This illustrates the importance of tailoring this function to the dataset.

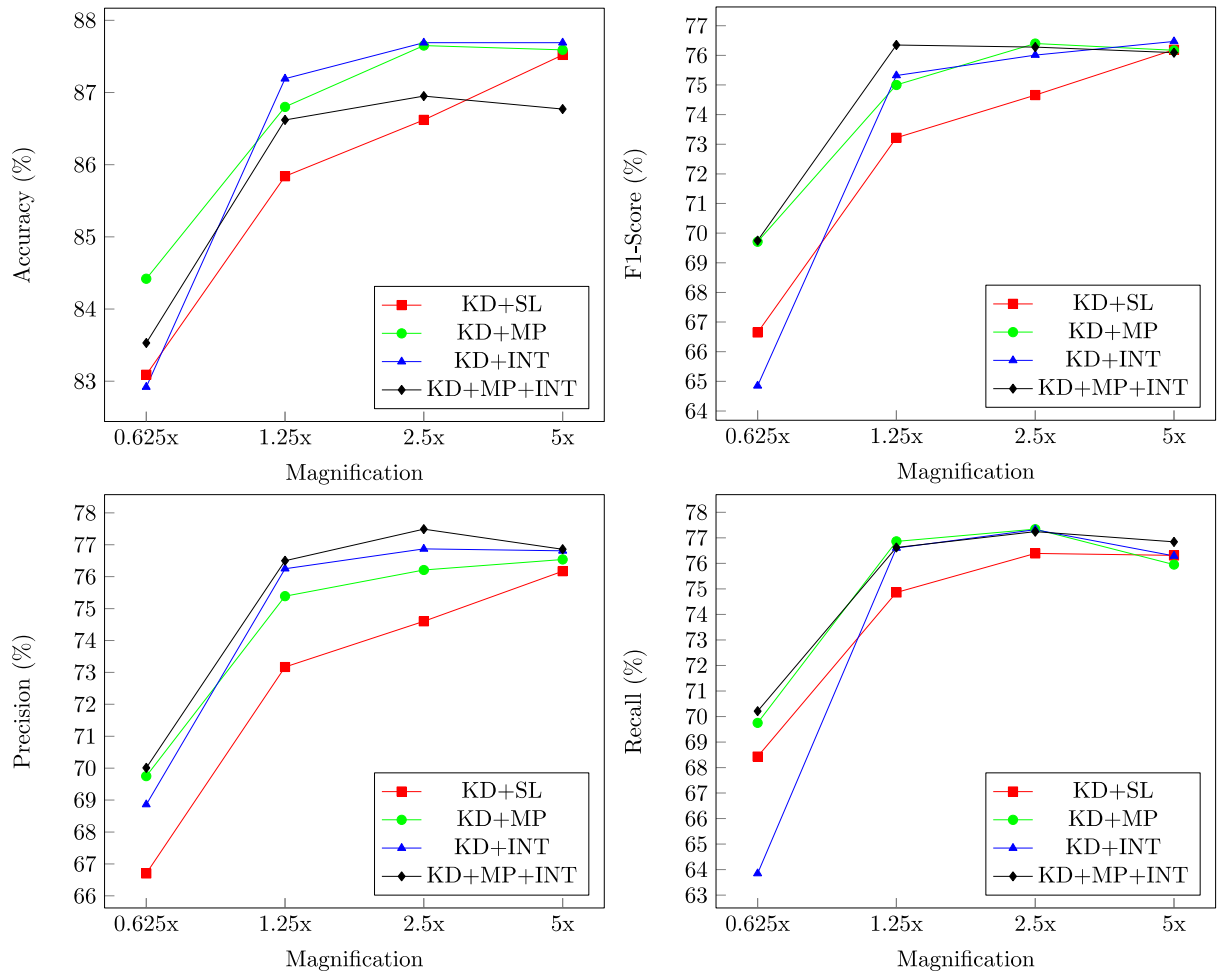

**Supplementary Figure 1.** Performance on a CD validation set for various resizing operations. The magnification corresponds to the training and validation data size.